\documentclass[preprint2]{aastex}
\usepackage{graphicx}

\shortauthors{Hopkins et al.}
\shorttitle{MicroJy sources in Abell 2877}

\begin{document}

\title{Microjansky radio sources in DC0107$-$46 (Abell 2877)}

\author{A. Hopkins\altaffilmark{1}} 

\affil{Department of Physics and Astronomy, University of Pittsburgh, 
  3941 O'Hara Street, Pittsburgh, PA 15260, USA}

\and

\author{A. Georgakakis}

\affil{School of Physics and Astronomy, University of Birmingham,
 Edgbaston, Birmingham B15 2TT, UK}

\and

\author{L. Cram}

\affil{School of Physics, University of Sydney,
  NSW 2006, Australia}

\and

\author{J. Afonso and B. Mobasher}

\affil{Astrophysics Group, The Blackett Laboratory, Prince Consort Road,
 London SW7 2BZ, UK}

\altaffiltext{1}{Previously at the School of Physics, University of
Sydney, NSW 2006, Australia. Current email address:
ahopkins@phyast.pitt.edu}

\begin{abstract}
The cluster DC0107$-$46 (Abell 2877) lies within the Phoenix Deep
Survey, made at 1.4\,GHz with the Australia Telescope Compact Array. Of
89 known optical cluster members, 70 lie within the radio survey area.
Of these 70 galaxies, 15 (21\%) are detected, with luminosities as faint as
$1.0 \times 10^{20}$\,W\,Hz$^{-1}$. Spectroscopic observations are
available for 14/15 of the radio-detected cluster galaxies. Six
galaxies show only absorption features and are typical low-luminosity
AGN radio sources. One galaxy hosts a Seyfert 2 nucleus, two are
star-forming galaxies, and the remaining five may be star-forming
galaxies, AGNs, or both.
\end{abstract}


\section{Introduction}
\label{int}

While ``classical'' radio galaxies powered by an active galactic
nucleus (AGN) dominate 1.4\,GHz source counts above $\approx $1\,mJy,
deeper radio surveys reveal a ``new'' population of radio galaxies at
sub-millijansky levels. Spectroscopy and multicolour photometry
\citep{Wind:85,Thu:87,Ben:93,Wind:94,Ham:95,Age:99} show that the
faint radio population comprises an increasing proportion of star
forming galaxies and a decreasing proportion of AGN-powered
galaxies. The evolutionary status of the star-forming galaxies, and
the role of interactions and mergers in the population are only
partially understood. This paper presents one of the first studies of
the sub-mJy population in an Abell cluster \citep[see also][]{Mof:89}.

Radio surveys of clusters of galaxies typically detect only a few
sources that are true cluster members. For example, \citet{Rob:90}
found that the detection rate of bright radio sources (S$_{408\,{\rm
MHz}} > 700$\,mJy) in Abell clusters is about 5\% when corrected for
chance coincidences. Any such detections are intrinsically luminous,
AGN-powered radio galaxies. At fainter flux density limits,
(S$_{1.4\,{\rm GHz}} > 10$\,mJy), \citet{LO:95} detected at least one
radio source in just under 40\% of a distance-limited cluster sample
($z \leq 0.09$), with a mean number of 1.4 radio sources per
cluster. Similar studies have been reported by \citet{Stoc:99} and
\citet{Zhao:89}. The sources detected in these deep surveys
are again almost all AGN-powered, with radio luminosities greater than
$10^{23}$\,W\,Hz$^{-1}$. Late-type galaxies in clusters are sometimes
detected \citep[e.g.,][]{AO:95}, but less frequently due to their low
radio luminosity and low space density.

The importance of excising cluster radio source contributions in
measurements of the Sunyaev-Zeldovich effect has prompted a few deep
surveys of selected clusters \citep[e.g.,][]{Mof:89,Her:95,Mye:97}.
However, these studies have not intended to explore the astrophysical
properties of the detected sources, and have not done so.

This paper discusses the cluster radio sources found in a deep 1.4\,GHz
survey (the {\em Phoenix Deep Survey}, PDS) which covers part of
DC0107$-$46 (Abell 2877, see \citeauthor{Dres:80b} \citeyear{Dres:80b} and
\citeauthor{ACO:89} \citeyear{ACO:89}; Appendix~\ref{app} summarises the
main properties of this cluster). The survey region was chosen to avoid if
possible sources brighter then $S_{5\,{\rm GHz}} \approx 30$\,mJy, and it
contains none of the apparently bright, intrinsically luminous, AGN-powered
sources described above. We show, however, that the sub-mJy radio source
population in the cluster is dominated by low-power (L$_{1.4} <
10^{22}$\,W\,Hz$^{-1}$) sources which are still generally powered by AGNs.
We adopt $H_0=75\,$kms$^{-1}$Mpc$^{-1}$ and $q_0=0.5$ in the following
analysis and discussion.

\section{Observations}
\label{obs}

The {\em Phoenix Deep Survey} \citep[PDS,][ and references
therein]{Hop:99} covers a high-latitude region of low optical
obscuration and devoid of bright radio sources. Australia Telescope
Compact Array (ATCA) observations of the PDS field at 1.4\,GHz reach a
$5\,\sigma$ limiting flux density of $300\,\mu$Jy over a two degree
diameter field, and encompass a $50\arcmin$ diameter field having a
$5\,\sigma$ level ranging from $100\,\mu$Jy at the perimeter to
$50\,\mu$Jy at its most sensitive \citep{Hop:99,Hop:98}. A radio
source list has been compiled using the {\sc sfind} package of {\em Miriad}.

Optical observations of the PDS field have been undertaken using the
Anglo-Australian Telescope (AAT). CCD imaging includes R-band photometry
for almost the complete field and V-band for slightly more than half
\citep{Age:99}. Optical catalogues constructed from the CCD images using
{\sc focas} (\citeauthor{JT:81} \citeyear{JT:81}; see \citeauthor{Age:99}
\citeyear{Age:99} for details of the catalogue construction) are complete
to $R \approx 22.5$. Optical
identifications of the radio sources have been attempted by cross-matching
the radio and optical catalogues. Approximately 50\% of the radio sources
have candidate optical identifications. Optical spectra of over 200 of the
optical candidates have been taken using the 2 degree field facility (2dF)
of the AAT. These have been analysed to yield redshifts and, in many cases,
spectroscopic classifications of the host galaxy \citep{Age:99}.

The 2 degree PDS field overlaps part of A2877. While almost all of the PDS
galaxies lie beyond the cluster, the overlap nevertheless provides an
opportunity to explore the sub-mJy population in the cluster itself. This
exploration is assisted by a study of galaxy cluster dynamics by
\citet{Mal:92} reporting redshifts for 125 galaxies which were candidates
for membership of A2877. Of these 125 objects, spectroscopy by
\citet{Mal:92} showed that 2 are foreground galaxies and a further 34 have
radial velocities outside the $3\,\sigma$ clipped sample or lying in a
sheet behind the cluster. This leaves 89 spectroscopically confirmed
cluster galaxies.

Radio counterparts of these A2877 members have been sought by
cross-matching the optical sources of \citet{Mal:92} with the PDS radio
catalogue. The region of the PDS overlaps about 75\% of A2877 and contains
70 of the 89 cluster galaxies within the radio survey area. Of these 70
galaxies, 15 (21\%) are detected at 1.4\,GHz. There are 6 other radio
detections identified with galaxies excluded from the cluster by
\citet{Mal:92}: 2 from the background sheet (velocities between 9000 and
10000\,kms$^{-1}$) and 4 at higher redshifts ($z=0.058,0.076,0.089,0.133$).
The 1.4\,GHz flux density for the radio-detected cluster galaxies ranges
from 0.1\,mJy to 14\,mJy. The cD galaxy is the fifth brightest source at
1.6\,mJy. Of the 15 PDS/A2877 sources, 14 have 2dF spectra.
Table~\ref{gals} presents radio flux densities, optical magnitudes,
and morphologies for the 15 radio-detected galaxies. The
redshifts and luminosities are also shown, and H$\alpha$ luminosities are
given for the galaxies with H$\alpha$ emission.

\section{Results}
\label{res}

Figure~\ref{fig1} shows radio contours overlayed on the AAT CCD images for
the 15 radio-detected cluster galaxies. These images were produced using
the {\sc kview} application from the {\em Karma\/} software suite
\citep{Goo:95}. Figure~\ref{spectra} shows spectra of the 8 radio-detected
cluster galaxies with emission features. It is important to note that
the projected area of the 2dF fibres is approximately 1\,kpc at the
distance of A2877, so that these spectra are from the nuclear regions only.
The $6''$ beam of the ATCA 1.4\,GHz observations corresponds approximately
to 3\,kpc at the cluster distance.

Notes on individual cluster objects:

  PDF~J010837.1$-$454819: The radio emission in this S0 galaxy is extended.
The spectrum shows H$\alpha$, [N{\sc ii}]($\lambda 658$\,nm) and
[S{\sc ii}]($\lambda\lambda 672, 673$\,nm) emission, but no H$\beta$,
[O{\sc iii}]($\lambda 501$\,nm) or [O{\sc i}]($\lambda 630$\,nm).
The [N{\sc ii}]/H$\alpha$
and [S{\sc ii}]/H$\alpha$ equivalent width (EW) ratios are 0.5 and 0.4,
respectively. In spectral diagnostic diagrams, \citep[e.g.,][]{VO:87}, such
ratios place the galaxy on the locus separating star forming galaxies from
Seyfert 2s and LINERs. Both AGN and star formation processes may be
present.

PDF~J010904.5$-$454624: Also catalogued as ESO~243$-$~G~045, this galaxy has
an E/S0 morphology and unresolved radio emission centred on the optical
nucleus. The spectrum shows no emission and has absorption lines
characteristic of an evolved stellar population. The radio emission is
likely to be due solely to an AGN.

  PDF~J010937.8$-$455350: The radio contours in this S0 galaxy are
displaced from the optical nucleus, perhaps indicating a radio lobe.
No emission lines are seen in this galaxy's spectrum, consistent with
the radio source being excited by an AGN.

  PDF~J010946.5$-$454657: This galaxy has an S0 morphology, with radio
contours centred on the optical nucleus. The spectrum shows emission lines
which unambiguously identify it as a Seyfert 2 or LINER in spectral diagnostic
diagrams. The absence of strong [O{\sc i}] emission and the presence of strong
[O{\sc iii}] emission suggest a Seyfert 2 interpretation \citep{Heck:80}. The
galaxy was also detected by \citet{CR:97} with an emission-line spectrum
(their Figure~10a, spectrum 43b).

  PDF~J010947.9$-$455125: This S0 galaxy has radio contours centred
on the optical nucleus. The spectrum shows H$\alpha$, [N{\sc ii}] and
[S{\sc ii}] emission, while H$\beta$ is present in absorption, and no
[O{\sc iii}] is detected. The [N{\sc ii}]/H$\alpha$ and [S{\sc ii}]/H$\alpha$
EW ratios of 0.6 and 0.3, respectively, give conflicting suggestions as
to the nature of the activity. The relative strength of the
[N{\sc ii}]/H$\alpha$ emission is suggestive of a Seyfert 2 nucleus, but
[S{\sc ii}] and [O{\sc iii}] are rather weak. Both AGN and star-formation
processes may be present. The presence of H$\beta$ in absorption
suggests that this is a post-starburst galaxy, possibly with a current
AGN \citep[e.g.,][]{Smail:99}.

  PDF~J010955.5$-$455552: This is IC 1633, the cluster cD galaxy. The
radio contours are centred on the optical nucleus. An extension to the
north-east is possible evidence of a radio lobe. The spectrum shows no
emission features, and has absorption lines characteristic of an
evolved stellar population.

  PDF~J011018.0$-$455556: This S0 galaxy has extended radio contours
centred on the optical nucleus. The outer parts of the optical object
have discernible structure, perhaps due to dust. A nearby, brighter radio
source is associated with a faint ($R>22.5$) optical object. This is
most likely a background source, but could possibly be a dwarf
companion showing interaction-induced star formation. The spectrum
shows no identifiable emission features, and evolved stellar
population absorption features are clearly detected.

  PDF~J011019.4$-$455113: This S0 galaxy is the strongest radio source
in the sample. The radio contours are displaced significantly from the
optical nucleus. No spectrum is available. The radio source could be (1) a
coincidentally located background source; (2) nuclear emission
displaced due to astrometric errors; (3) strong, localised disk star
formation; (4) a radio lobe; or (5) a radio supernova associated with
the galaxy. Case (1) is unlikely, as there is only a 0.2\% chance of
an accidental coincidence between optical and radio sources of this
magnitude. The close agreement between the positions of the weak radio
source to the southwest and its faint optical counterpart argue
against a large astrometric error. The radio emission is of a luminosity
about an order of magnitude greater than that due to the extreme
Type II radio supernova RSN 1986J \citep{Wei:90,SLL:98}. Hence, if
a supernova origin for the radio emission is correct, it is unlikely
to be due to a single object. Perhaps several radio supernovae are
present in a localised burst of star formation. While planned optical
spectroscopy will detect any star formation activity, a single displaced
radio lobe remains another possible interpretation.

  PDF~J011027.6$-$460428: Also catalogued as ESO~243$-$~G~049, this S0
galaxy has radio contours centred on the optical nucleus. The
spectrum shows evolved stellar population absorption
features. An AGN origin for the radio emission is likely.

  PDF~J011029.4$-$461027: Dressler's classification is Sb(p). This
face-on flocculent spiral shows knots of optical emission towards the
nucleus. The radio contours are extended but centred on the optical
nucleus. The spectrum has emission lines which indicate the galaxy
has strong H{\sc ii} regions. The IRAS 60$\mu$ flux density is 0.455\,Jy,
giving $S_{60\mu}/S_{1.4\,{\rm GHz}}=96.0$, consistent with the
well-established radio/FIR correlation for spiral galaxies. Star
formation is the likely cause of the radio emission in this galaxy.
The calibration of radio luminosity to SFR given in \citet{Cram:98}
implies this galaxy has SFR$_{1.4}=1.3\,M_{\odot}$yr$^{-1}$. This
is significantly higher than the SFR derived from the H$\alpha$ luminosity,
(again using the relations given by \citeauthor{Cram:98} \citeyear{Cram:98}),
SFR$_{H\alpha}=0.08\,M_{\odot}$yr$^{-1}$.

  PDF~J011047.2$-$454701: This S0 galaxy has six apparent dwarf companions in
the optical image. The radio contours are displaced from the optical
nucleus, and two apparent companions show radio emission. The spectrum
shows an evolved stellar population and no hint of emission. This suggests
that the offset radio source could be the radio lobe of an AGN. This group
of radio and optical sources is quite unusual and invites further study.

  PDF~J011055.8$-$453920: This galaxy is present but unclassified in
Dressler's catalogue. The optical isophotes hint at a three-armed spiral or
the presence of a second nucleus. The radio contours, centred on the peak
of the optical emission, show a slight extension in the direction of the
optical asymmetry. The spectrum shows emission features which clearly
define the galaxy as an H{\sc ii}-region type in spectral diagnostic diagrams.
The radio emission is probably due to star formation. If so, this is the
only galaxy in the cluster whose unusual optical morphology gives possible
evidence of merger-induced star formation. The implied SFRs from
1.4\,GHz and H$\alpha$ luminosities are 0.08 and $0.28\,M_{\odot}$yr$^{-1}$
respectively. Note that the H$\alpha$ luminosity in this case has been
derived from the equivalent width of the emission line.

  PDF~J011119.1$-$455554: This galaxy is ESO~243$-$~G~051. Dressler's
classification is Sb. The radio contours are centred close to the
optical nucleus and show extension tracing the optical emission. The
spectrum shows absorption features with weak H$\alpha$ and [N{\sc ii}]
emission, with [N{\sc ii}] stronger and indicative of an AGN. However, the
60$\mu$ flux density is 0.598\,Jy, giving $S_{60\mu}/S_{1.4\,{\rm
GHz}}=90.0$, consistent with the well-established radio/FIR
correlation for star-forming galaxies. Both AGN and star-formation
processes may be present. The radio and H$\alpha$ luminosities can be
used in this case to provide an upper limit to the SFR, giving
SFR$_{1.4}<3.5\,M_{\odot}$yr$^{-1}$ and
SFR$_{H\alpha}<0.009\,M_{\odot}$yr$^{-1}$. Again, the SFR from H$\alpha$ is
much lower than the radio estimate.

  PDF~J011153.5$-$455847: This galaxy has early-type optical morphology.
The radio source is displaced from the optical nucleus. The spectrum
shows emission features and the galaxy's location in spectral
diagnostic diagrams is ambiguous. The radio emission may be coming from
both star formation and nuclear activity. The large nearby edge-on spiral
galaxy is unrelated, being at a redshift of 0.089, but there is a faint
apparent optical companion lying close to the direction of displacement of
the radio emission.

  PDF~J011219.3$-$455322: Catalogued as ESO~244$-$~G~001, this galaxy
has an Scd morphology, and shows knotty or dusty features. The radio
emission is asymmetric with respect to the optical nucleus, and
extended southward where no optical emission is present. The apparent
optical companions show no radio emission, and a low-level radio
companion is also present with no optical counterpart. The spectrum
shows emission features, and has an ambiguous classification between
diagnostic diagrams. This galaxy may be hosting both star formation
and AGN activity. The implied upper limits to the SFRs derived from radio and
H$\alpha$ luminosities are 0.3 and $0.06\,M_{\odot}$yr$^{-1}$ respectively,
again with SFR$_{H\alpha}$ much lower than SFR$_{1.4}$.
 
\section{Discussion}
\label{disc}

Of the 70 spectroscopically confirmed members of A2877 overlapping the
PDS, $64$\% have early-type (E/S0) morphology. This
is consistent with the 63\% quoted by \citet{CR:97}, who give
the proportions for the total of both Dressler clusters DC0107$-$46
and DC0103$-$47. Of the 15 galaxies with radio detections, 3 have Sb
or Scd morphology. A fourth, unclassified galaxy is likely to be a
spiral or peculiar spiral. The remainder are S0 galaxies apart from
the cD, one E/S0 and one unclassified but early-type galaxy. There is
no obvious tendency for radio detections to favour a particular Hubble
type in this cluster.

Our survey is unusually deep for a radio survey of an Abell cluster, and it
detects a relatively large fraction ($\approx 20$ \%) of the optical
cluster members. However, our results are consistent with the known
properties of early-type galaxies in the field. \citet{Sad:89} studied
5\,GHz emission from nearby optically selected early type (E/S0) galaxies.
They found that 20--30\% of their sample (with optical luminosities similar
to those probed here) show radio emission with luminosities similar to
those of the sources in the present sample.

\citet{Sad:89} (hereafter SJK) find that the probability distribution of
radio luminosity is roughly uniform across the range of luminosities found
in our study ($S_{1.4} \approx 10^{20}$ -- $10^{22}$\,W\,Hz$^{-1}$), with
some tendency for a smaller fraction at higher luminosities. This is also
consistent with our study, as is demonstrated in Figure~\ref{flf1}
comparing the fractional bivariate luminosity function (FBLF) of SJK with
that of the present sample. The FBLF estimates the distribution of radio
luminosities ($L_{1.4}$) for galaxies in a given optical magnitude ($M_B$)
range. The construction of the FBLF for the cluster sample used 31 cluster
galaxies which fall in the luminosity ranges of Figure~\ref{flf1}
(with radio detections or upper limits from the PDS and $B_J$
magnitudes from the COSMOS database, \citeauthor{Drink:95} \citeyear{Drink:95};
\citeauthor{Yen:92} \citeyear{Yen:92}).
To improve the statistics of
this calculation, the galaxies with 1.4\,GHz upper limits were distributed
uniformly throughout the bins they could possibly occupy, rather than using
a more sophisticated maximum-likelihood method \citep[e.g.,][]{Av:80}. This
simple approach still gives consistent results, however, with the the
method described by \citet{Av:80}, where there are sufficient cluster
detections for its use.

There is good agreement between the samples in the two brighter magnitude
bins where the range of radio luminosities overlaps, and the apparent
excess for the cluster sample in the fainter bin is still consistent with
the data of SJK. Although no morphological distinction was made when
compiling the cluster sample for this analysis, differences resulting from
the inclusion of late-type cluster galaxies are small given that
the cluster is dominated by early-type galaxies. Within the statistical
uncertainty of our small sample, our cluster results are consistent with the
results of SJK.

The different distribution of Hubble-type between cluster and field
environments has long been known \citep[e.g.,][]{Dres:80a} and this does
influence the statistics of our cluster sample compared with the field.
According to \citet{Con:89}, for example, the space density of star-forming
{\it field} galaxies with $L_{1.4} \approx 10^{20}$ --
$10^{21}$\,W\,Hz$^{-1}$ is at least an order of magnitude larger than that
of early-type (AGN) galaxies. However, the galaxy population in A2877,
like other nearby clusters, is dominated by early types. The absence of a
statistically significant difference between our cluster sample and the SJK
early-type galaxy sample suggests that at least in A2877 the power of the
radio sources are not influenced by the cluster.

The few late-type galaxies positively classified in A2877 and detected in
the radio appear to have properties (such as FIR emission) analogous to
comparable field galaxies. For these objects it has been possible to
estimate star formation rates from the 1.4\,GHz and
H$\alpha$ luminosities. The estimates of H$\alpha$ luminosity derived from
EW measurements predict SFRs higher than the radio, and
those from flux calibrated spectra systematically lower. This may be the
result of two effects. First, the EW measurements yield an H$\alpha$
luminosity based on the R-band luminosity for the whole galaxy, which may
well overestimate the true value in low SFR galaxies \citep[c.f.][~their
Figure~1]{Cram:98}. Second, the flux calibrated spectra are sampling only
a small portion of these galaxies, and not necessarily the region
containing current star formation. The presence of any extinction would
serve to lower these luminosities even further. Hence it is unsurprising
that the SFRs implied from these luminosities are significantly lower than
that deduced from the radio (and the FIR). These results are not
inconsistent with the large scatter revealed by \citet{Cram:98}, although
the systematic underestimates are uncharacteristic of the trend they
observe at low star formation rates.

\section{Conclusion}
\label{conc}

We have detected 1.4\,GHz emission from 15 galaxies confirmed as members of
the cluster A2877. For 2 of these, the radio emission is likely to be due
to star formation processes, with star formation rates of 1.3 and
$0.08\,M_{\odot}$yr$^{-1}$. One source (PDF~J010946.5$-$454657) is probably
a Seyfert 2, and in a further 5 galaxies both AGN and star formation
processes may be contributing to the detected radio emission. Of the
remaining 7 galaxies, 6 show only absorption features in their spectra,
typical of evolved stellar populations, and one has no spectrum available.

Our deep survey has allowed us to probe radio luminosities at the distance
of the cluster to the same depth as that reached by \citet{Sad:89} in the
field, albeit with far fewer numbers in our sample. The fraction and
luminosity distribution of radio sources in the early-type galaxies in
A2877 is indistinguishable from the properties of field galaxies of the
same type. We do, however, find two examples in which the presence of
companions might play a role in the radio emission. Whether this is a
special characteristic of the cluster environment, however, we cannot say.

None of the radio-detected galaxies in the cluster show evidence for strong
morphological distortion. Interactions between galaxies and detectable
dwarf companions do not appear to be necessary to excite radio emission,
although in specific cases such interactions may still occur, with
PDF~J011047.2$-$454701 being one potential example.

\acknowledgements

This research has made use of the NASA/IPAC Extragalactic Database
(NED) which is operated by the Jet Propulsion Laboratory, California
Institute of Technology, under contract with the National Aeronautics
and Space Administration. The Digitized Sky Survey was produced at the
Space Telescope Science Institute under US Government grant NAG
W-2166. AMH and LEC acknowledge financial support from the Australian
Research Council and the Science Foundation for Physics within the
University of Sydney. JMA gratefully acknowledges support in the form
of a scholarship from Funda\c{c}\~{a}o para a Ci\^{e}ncia e a Tecnologia
through Programa Praxis XXI. The Australia Telescope is funded by the
Commonwealth of Australia for operation as a National Facility managed
by CSIRO.

\appendix
\section{Abell 2877}
\label{app}

Abell 2877 was catalogued by \citet{Dres:80b} as DC0107$-$46. It is the
richer half of a pair of clusters, DC0107$-$46 and DC0103$-$47
\citep[see also][]{CR:97}. Other names for the cluster include
AM~0107-461, SCL~018~NED05, APM~010740.1-461028 \citep{Dal:94}, and
EXSS~0107.6-4610. A2877 is Abell-type R (regular). It has a prominent
central cD galaxy and thus is Bautz-Morgan type I \citep{BM:70}. The
cluster lies at a redshift of $z=0.0241$ (distance class 2), and is
``poor'', having a richness class R=0 \citep{ACO:89}. A2877 shows evidence
of substructure \citep{Mal:92,Gir:97}, as do about a third of galaxy
clusters \citep{Gir:97}. The cluster is an X-ray source with a
temperature of $T_X \simeq 3.5\,$keV \citep{Dav:93}; it shows no
evidence for a cooling flow \citep[e.g.,][]{Whi:97}. \citet{Gir:98}
estimate a virial radius of $R_{\rm vir} \simeq 1.8\,h^{-1}\,$Mpc, a
velocity dispersion of $\sigma \simeq 900\,$kms$^{-1}$, and a
corrected virial mass of $M_{\rm CV} \simeq 5 \times
10^{14}\,h^{-1}M_{\odot}$.

\begin{deluxetable}{ccclcccc}
\tablewidth{0pt}
\tablecaption{Details of the radio-detected galaxies in A2877.
 \label{gals}}
\tablehead{
\colhead{Name} & \colhead{$S_{1.4}$} & \colhead{$R$} & \colhead{Hubble} & \colhead{$z$} & \colhead{$\log(P_{1.4})$} & \colhead{$M_R$} & \colhead{$\log(L_{H\alpha})$} \\
\colhead{} & \colhead{(mJy)} &\colhead{} & \colhead{type} &\colhead{} & \colhead{(W\,Hz$^{-1}$)} &\colhead{} & \colhead{(W)}
}
\startdata
  PDF~J010837.1$-$454819 & 0.61 & 15.85 & S0\tablenotemark{a} & 0.027 & 20.94 & $-$19.33 & 33.49\tablenotemark{b} \\
  PDF~J010904.5$-$454624 & 5.00 & 12.45 & E/S0 & 0.026 & 21.82 & $-$22.65 & \\
  PDF~J010937.8$-$455350 & 0.11 & 13.91 & S0 & 0.022 & 20.01 & $-$20.83 & \\
  PDF~J010946.5$-$454657 & 1.12 & 14.47 & S0 & 0.020 & 20.94 & $-$20.05 & 32.81 \\
  PDF~J010947.9$-$455125 & 0.35 & 14.96 & S0 & 0.020 & 20.43 & $-$19.56 & 33.27\tablenotemark{b} \\
  PDF~J010955.5$-$455552 & 1.59 & 11.09\tablenotemark{c} & D & 0.024 & 21.25 & $-$23.83 & \\
  PDF~J011018.0$-$455556 & 0.20 & 15.18 & S0\tablenotemark{a} & 0.028 & 20.49 & $-$20.08 & \\
  PDF~J011019.4$-$455113 & 14.3 & 14.23 & S0 & 0.024 & 22.20 & $-$20.69 & \\
  PDF~J011027.6$-$460428 & 0.18 & 13.52 & S0 & 0.023 & 20.27 & $-$21.31 & \\
  PDF~J011029.4$-$461027 & 4.69 & 14.30 & Sb(p) & 0.024 & 21.72 & $-$20.63 & 33.10 \\
  PDF~J011047.2$-$454701 & 0.10 & 14.62 & S0 & 0.025 & 20.09 & $-$20.40 & \\
  PDF~J011055.8$-$453920 & 0.28 & 15.74 & & 0.024 & 20.50 & $-$19.18 & 33.62\tablenotemark{b} \\
  PDF~J011119.1$-$455554 & 6.65 & 12.80\tablenotemark{c} & Sb & 0.022 & 21.80 & $-$21.93 & 32.12 \\
  PDF~J011153.5$-$455847 & 0.14 & 16.69 & & 0.025 & 20.24 & $-$18.33 & 33.19\tablenotemark{b} \\
  PDF~J011219.3$-$455322 & 0.87 & 15.28 & Scd & 0.027 & 21.09 & $-$19.90 & 32.92 \\
\enddata
\tablenotetext{a}{All the morphological types are
from \protect\citet{Dres:80b} apart from these two,
which have been taken from \protect\citet{CR:97}.}
\tablenotetext{b}{These H$\alpha$ luminosities were derived indirectly from
EWs \protect\citep{Age:99}.}
\tablenotetext{c}{These are Cousins $R_{25}$ magnitudes
measured by \protect\citet{LV:89}.}
\end{deluxetable}

\begin{figure*}
\centerline{\hfill
\rotatebox{0}{\includegraphics[width=5cm]{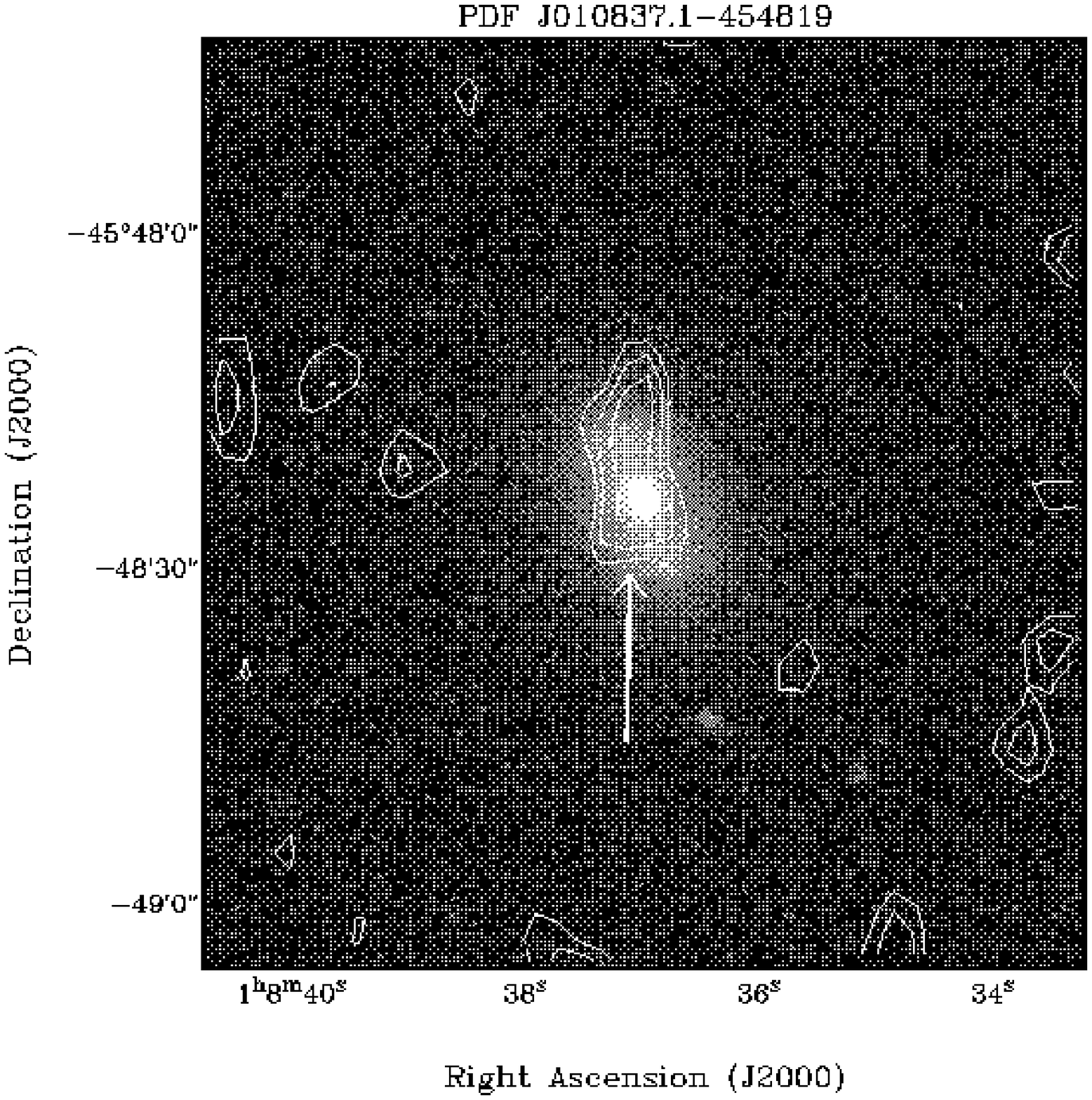}}
\hfill
\rotatebox{0}{\includegraphics[width=5cm]{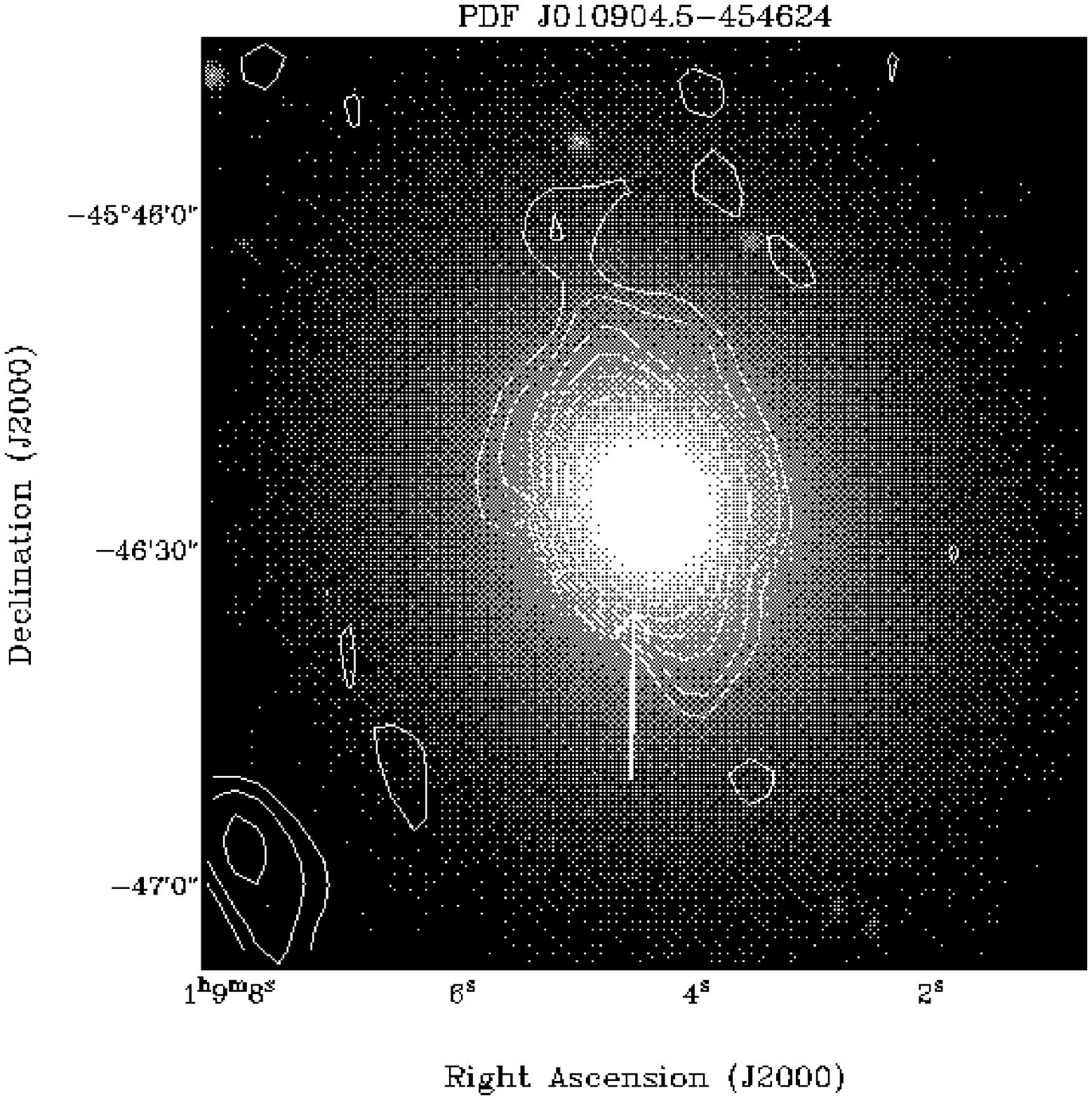}}
\hfill
\rotatebox{0}{\includegraphics[width=5cm]{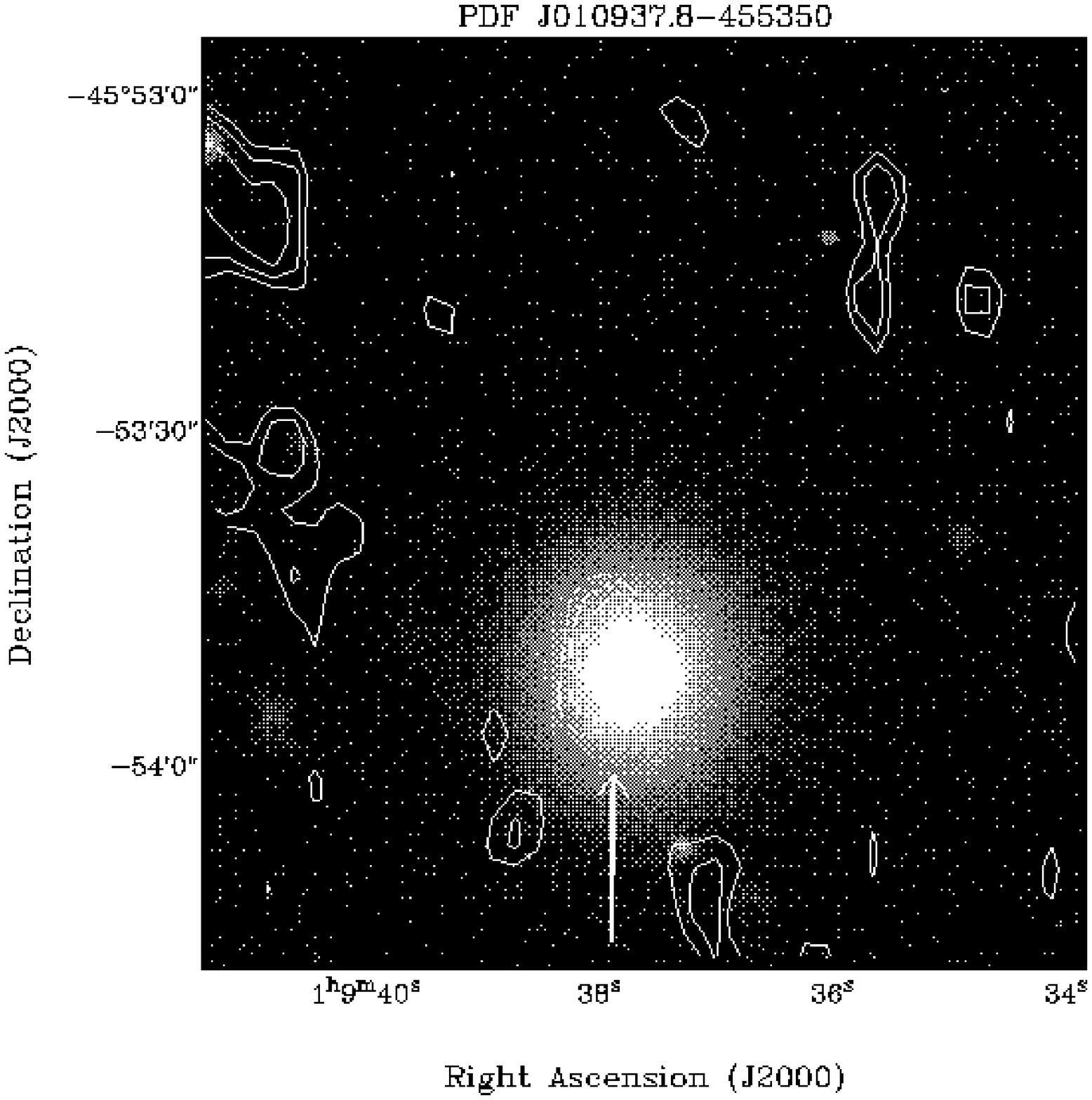}}}
\centerline{\hfill
\rotatebox{0}{\includegraphics[width=5cm]{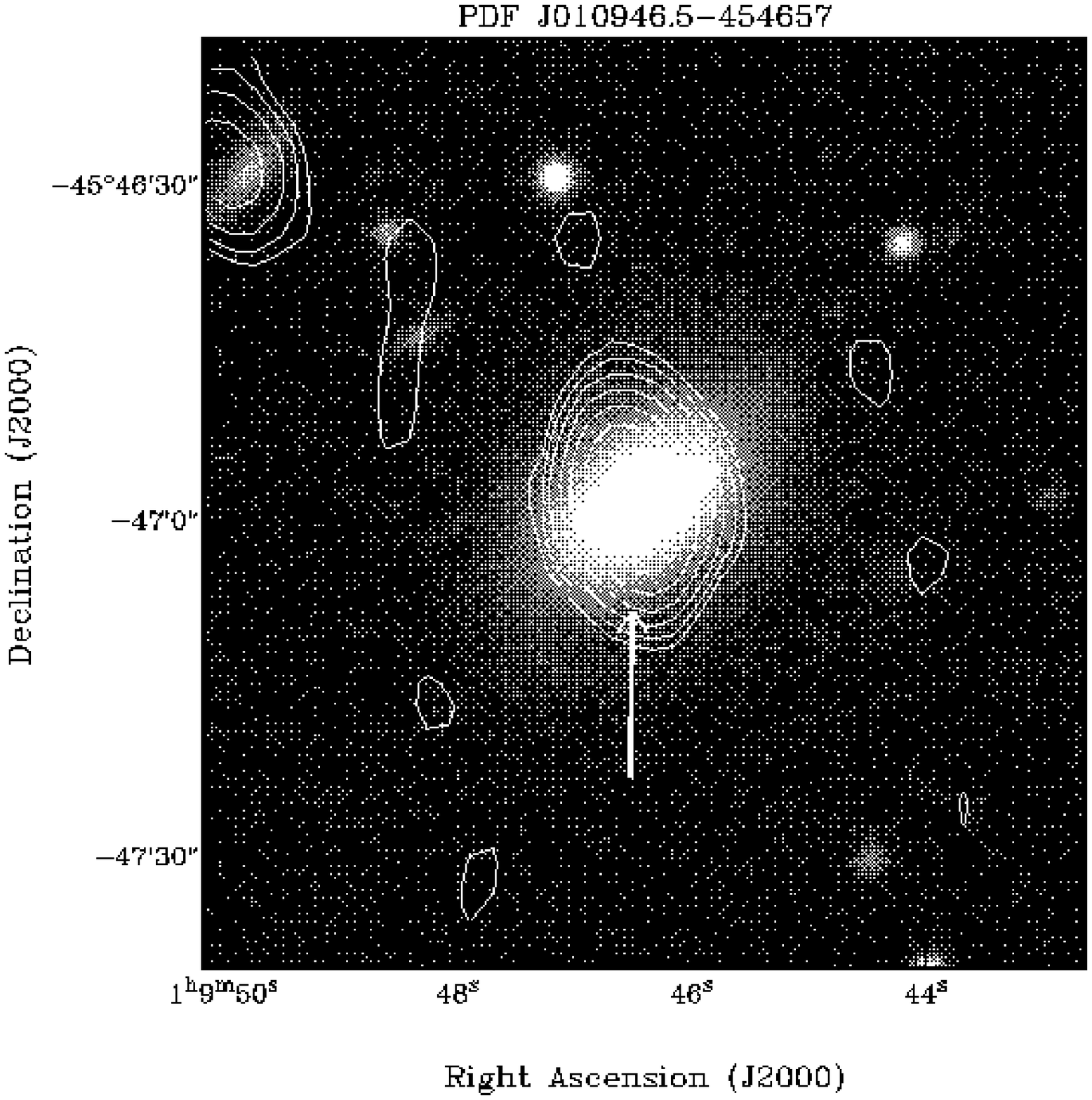}}
\hfill
\rotatebox{0}{\includegraphics[width=5cm]{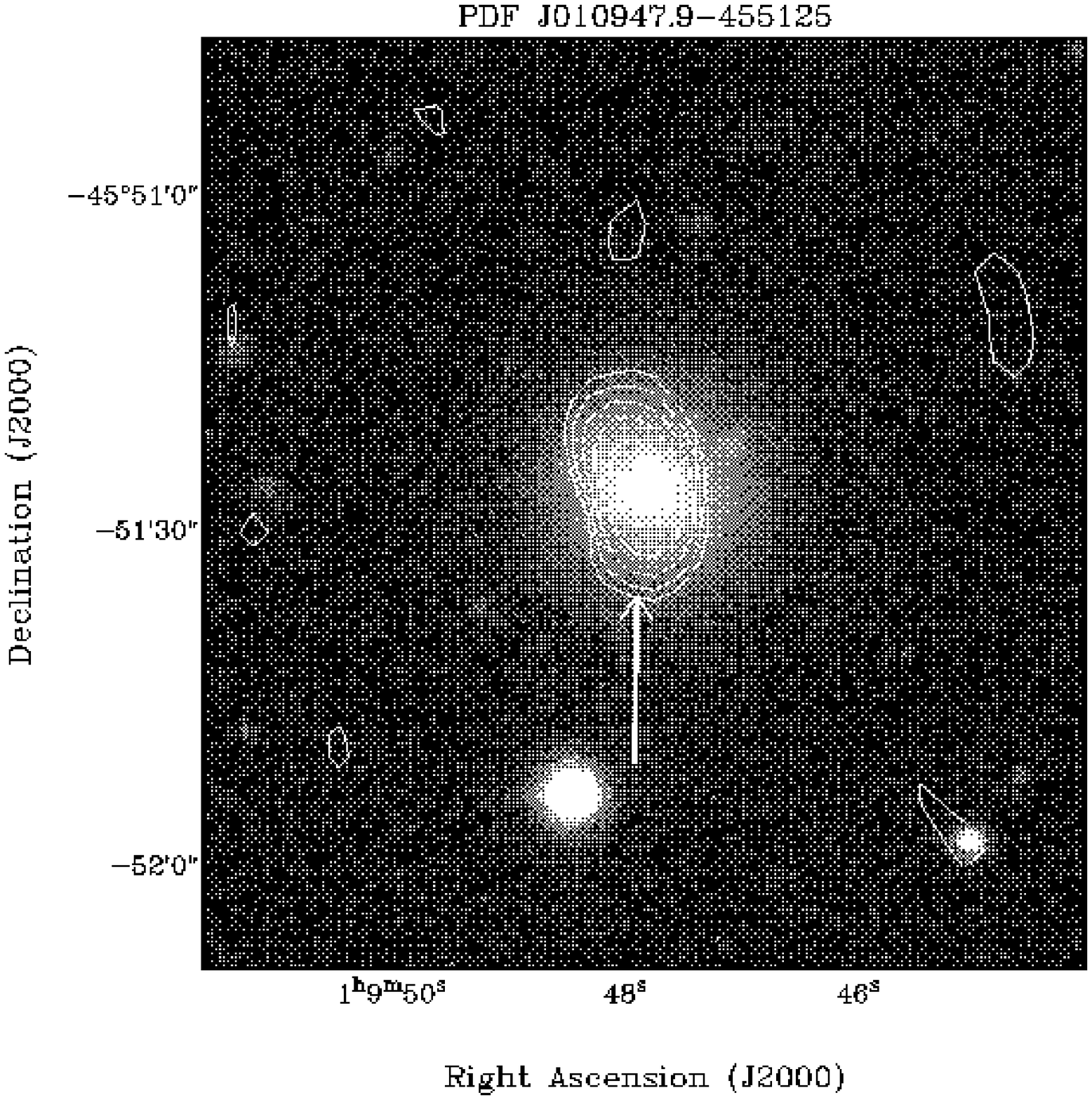}}
\hfill
\rotatebox{0}{\includegraphics[width=5cm]{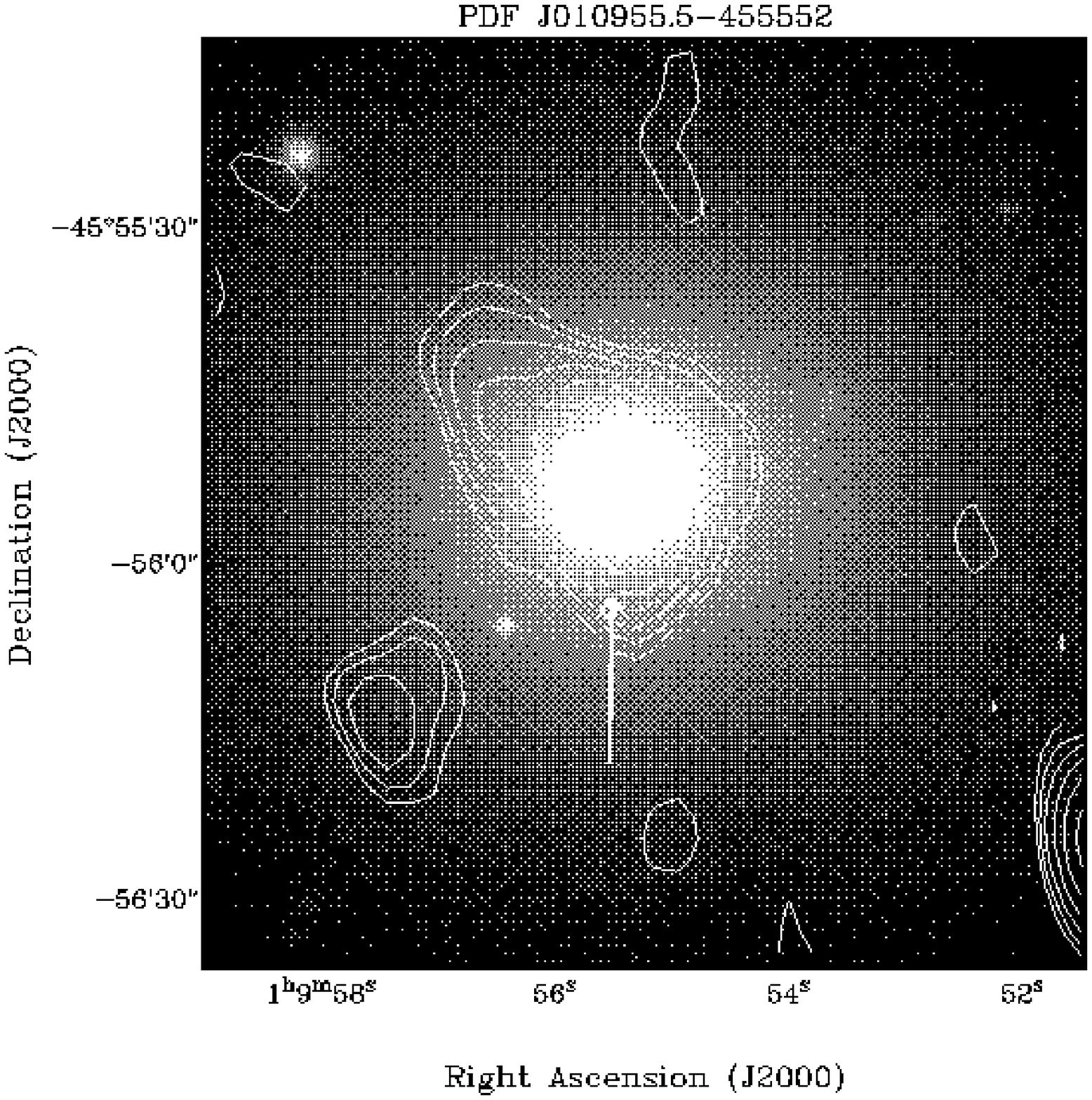}}}
\centerline{\hfill
\rotatebox{0}{\includegraphics[width=5cm]{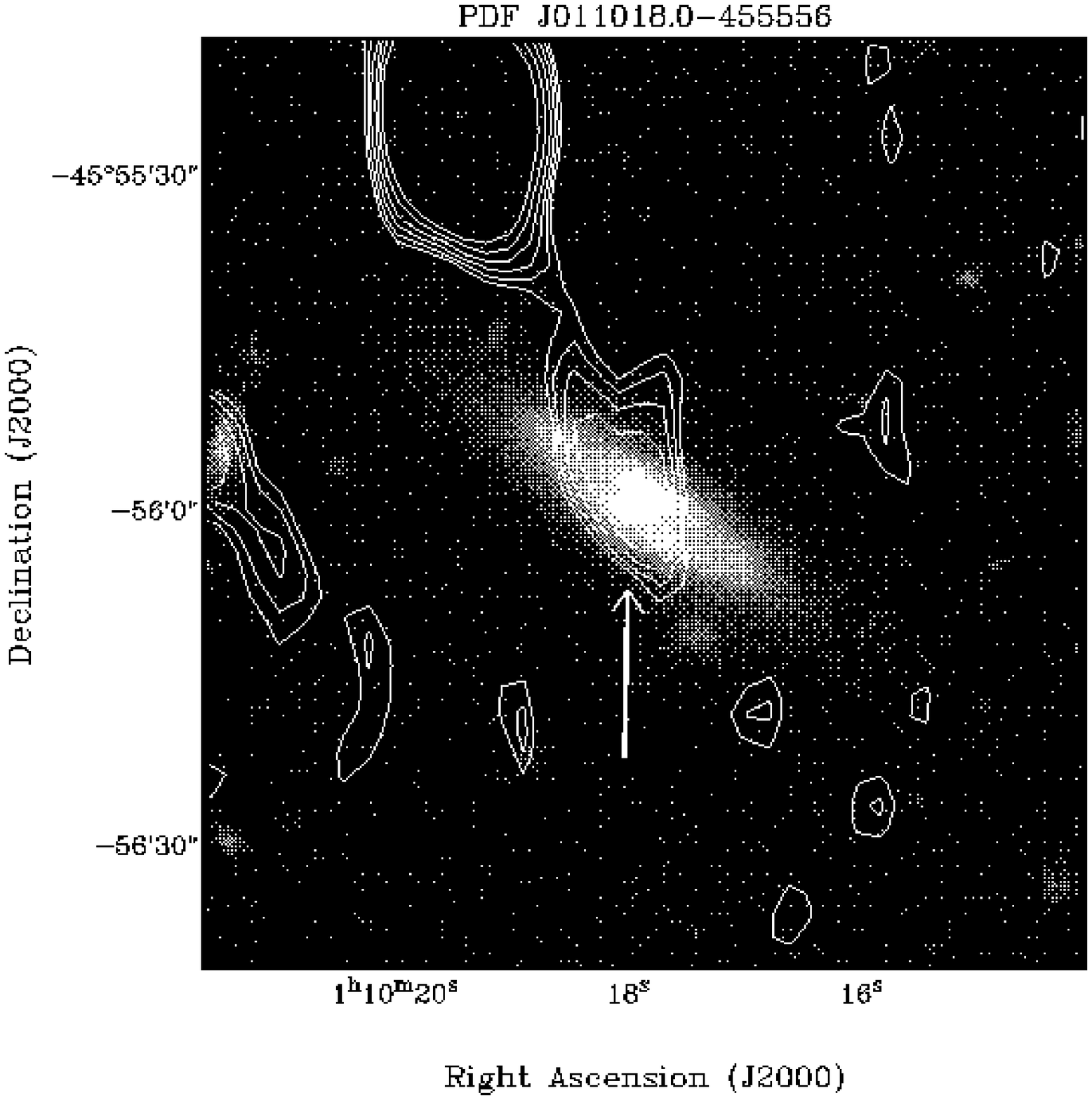}}
\hfill
\rotatebox{0}{\includegraphics[width=5cm]{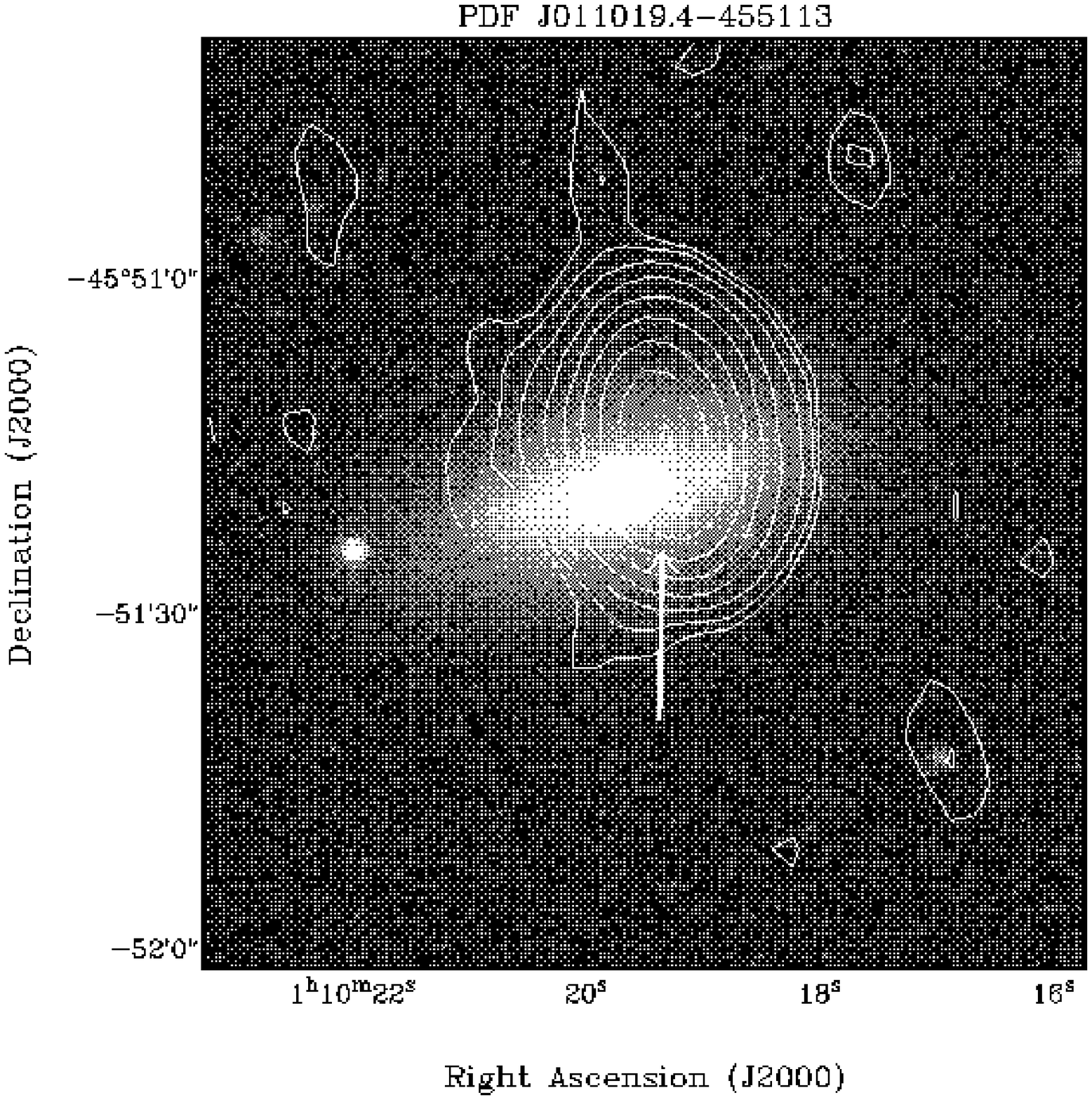}}
\hfill
\rotatebox{0}{\includegraphics[width=5cm]{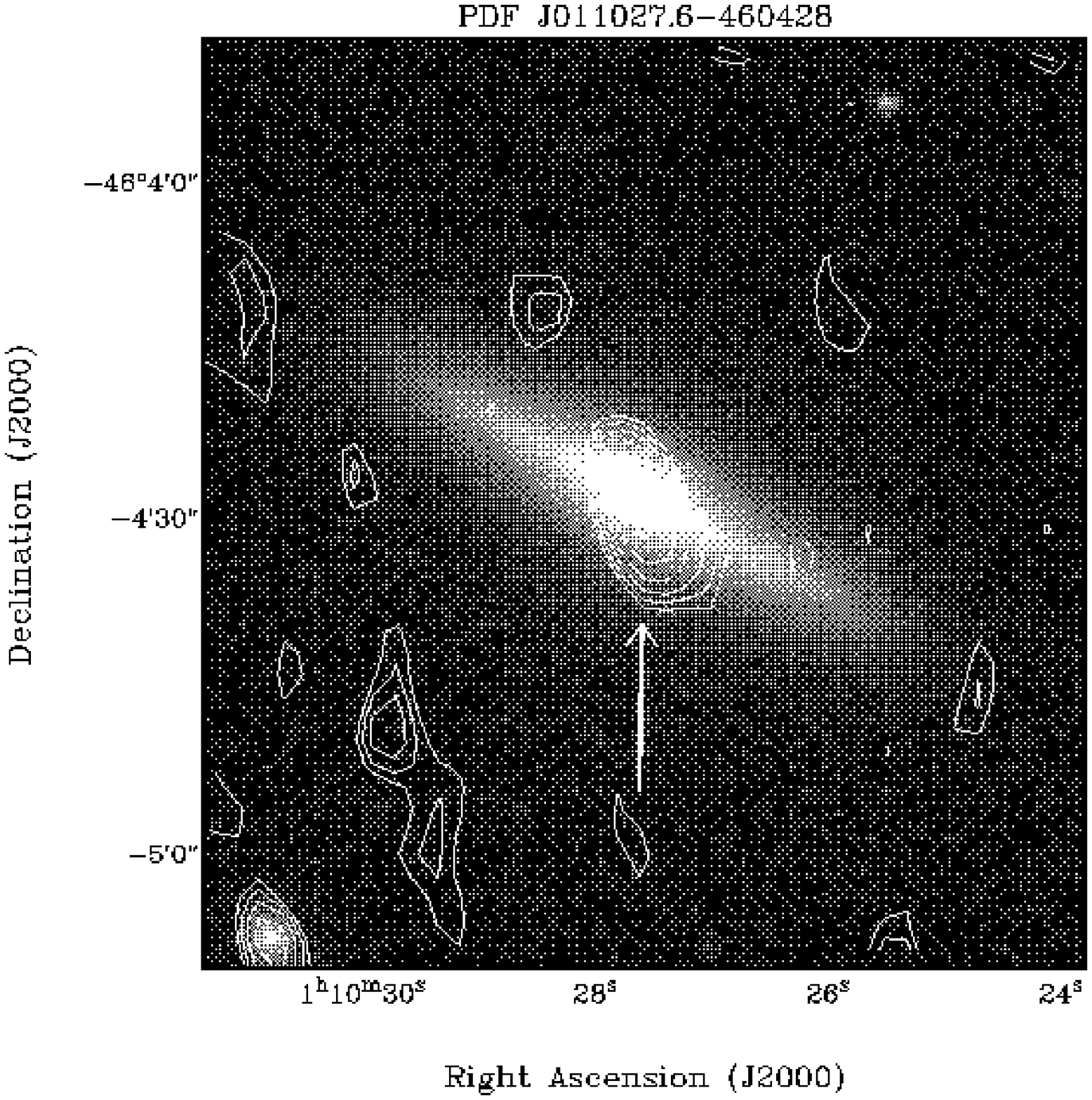}}}
\caption{Radio contours overlayed on the AAT CCD image for each of the
15 radio-detected cluster galaxies. An arrow indicates the radio
source in each frame. The object on the right in the second row is the
cD galaxy. \label{fig1}}
\end{figure*}

\addtocounter{figure}{-1}

\begin{figure*}
\centerline{\hfill
\rotatebox{0}{\includegraphics[width=5cm]{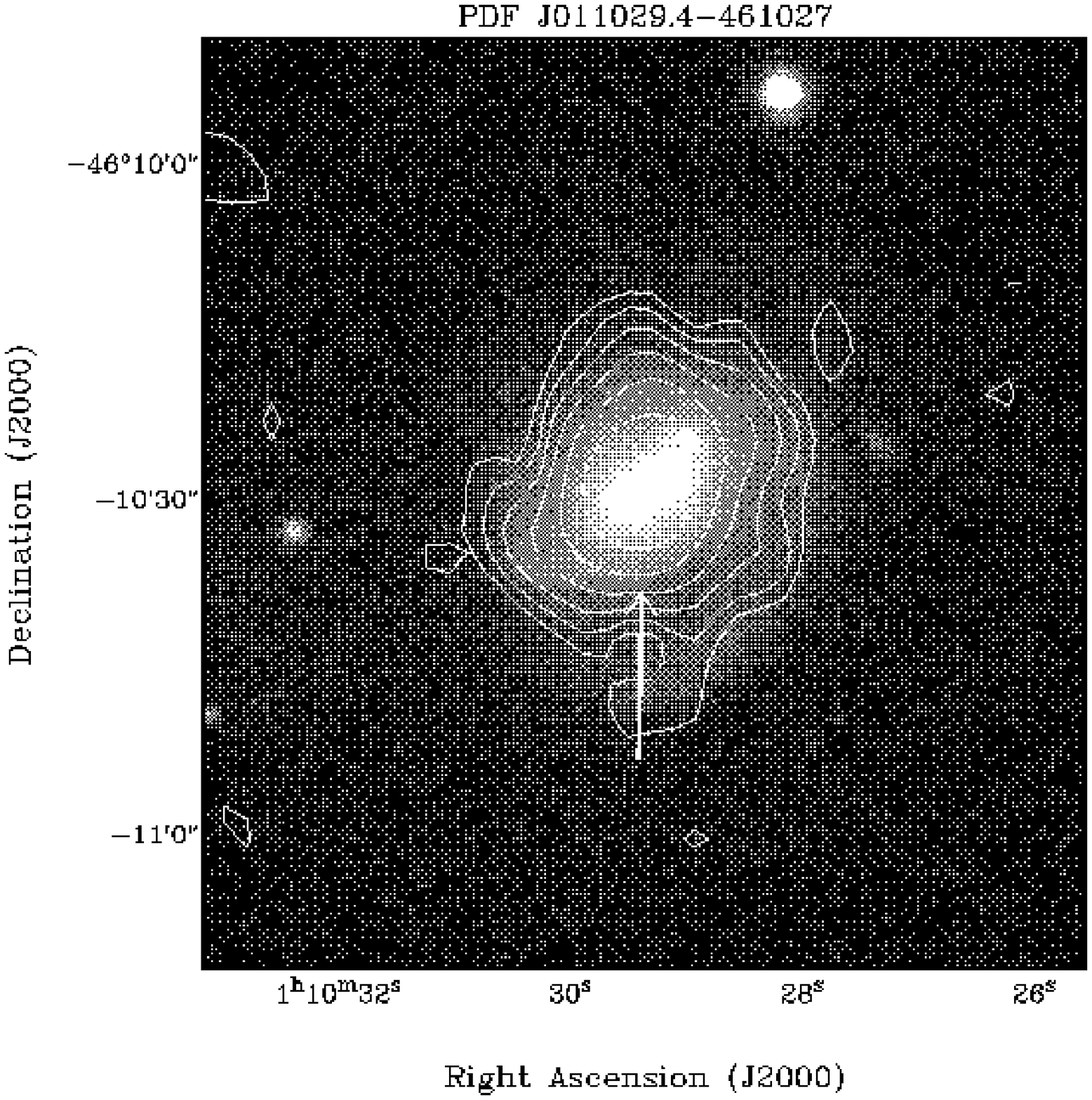}}
\hfill
\rotatebox{0}{\includegraphics[width=5cm]{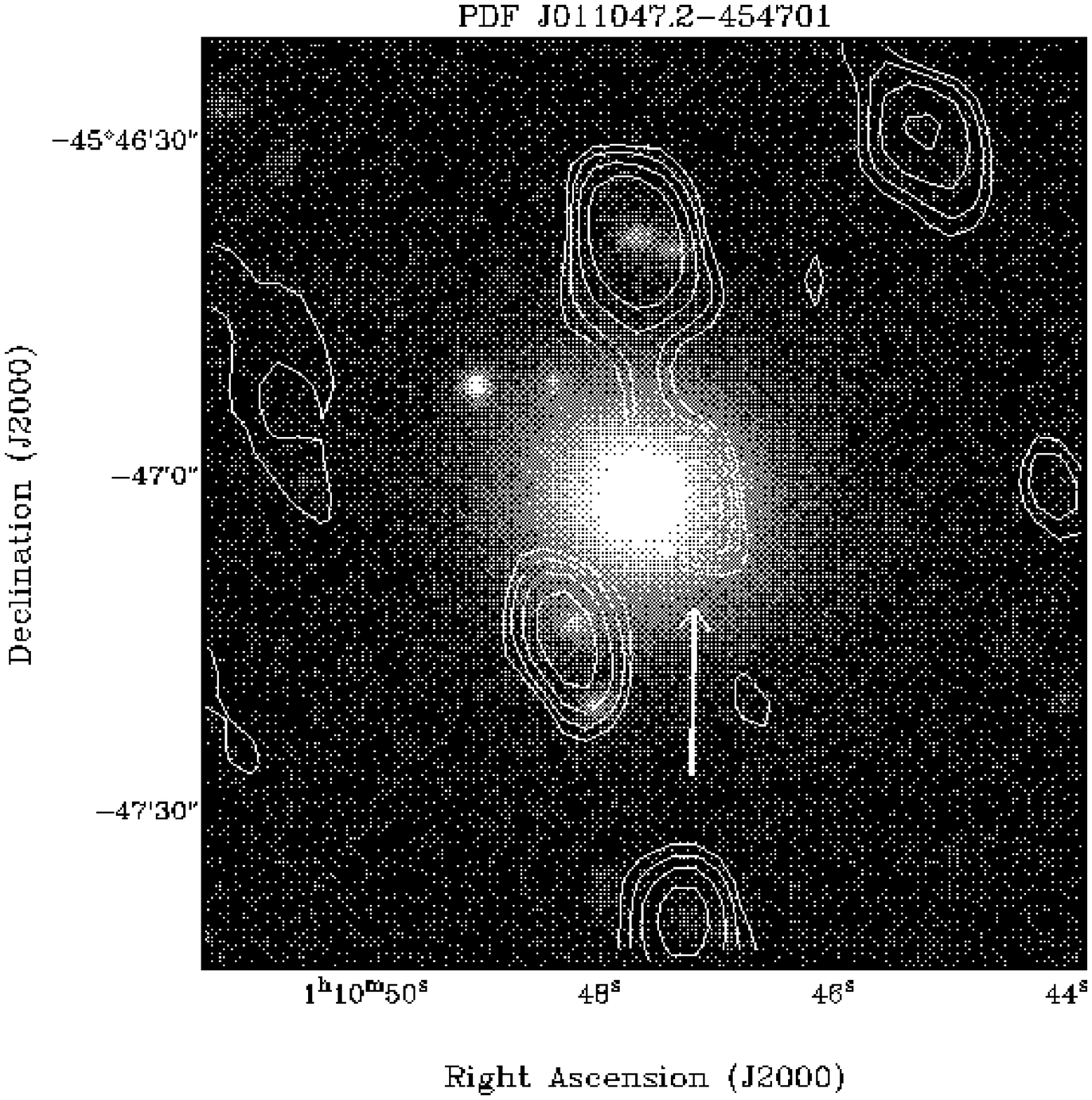}}
\hfill
\rotatebox{0}{\includegraphics[width=5cm]{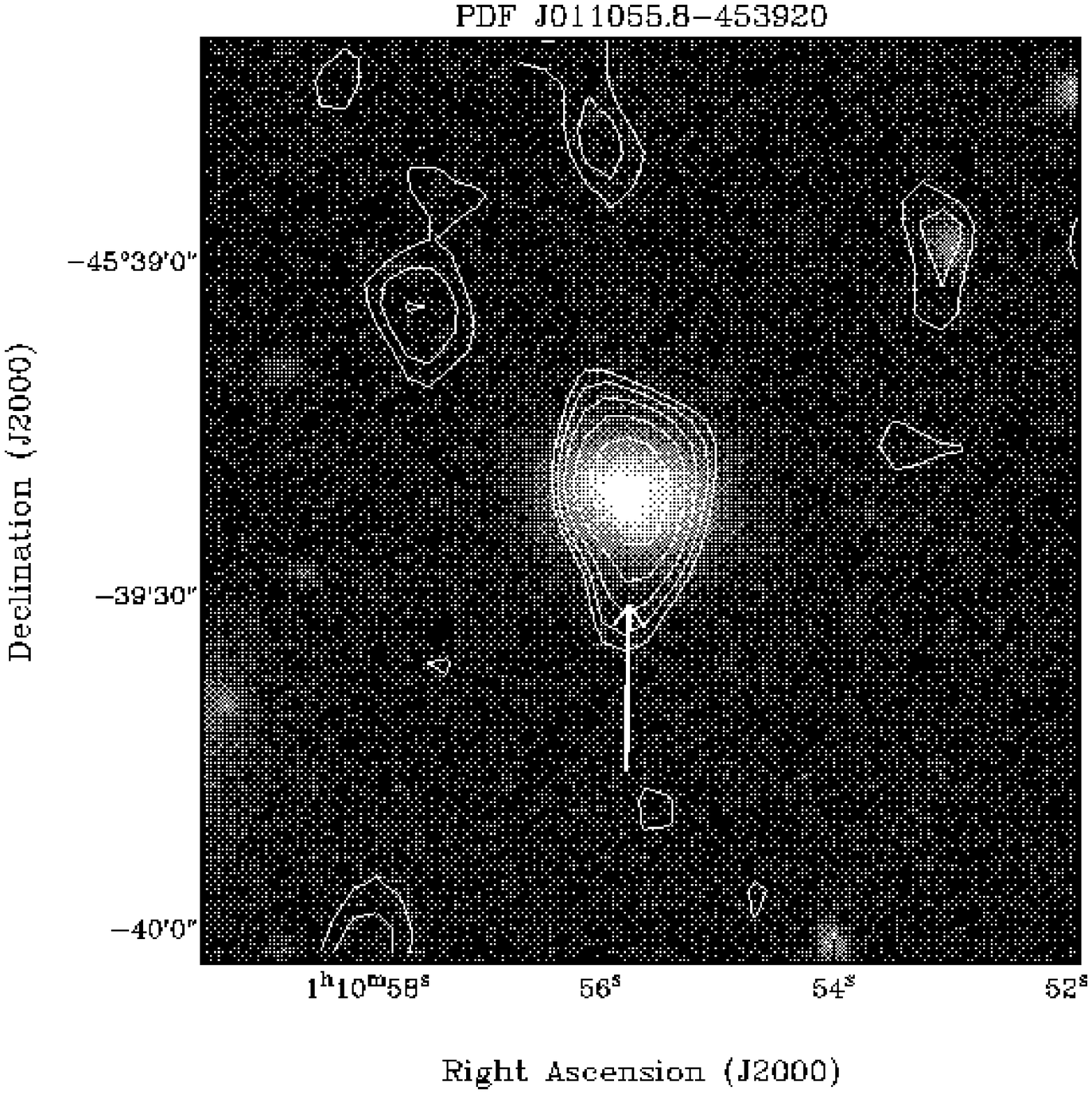}}}
\centerline{\hfill
\rotatebox{0}{\includegraphics[width=5cm]{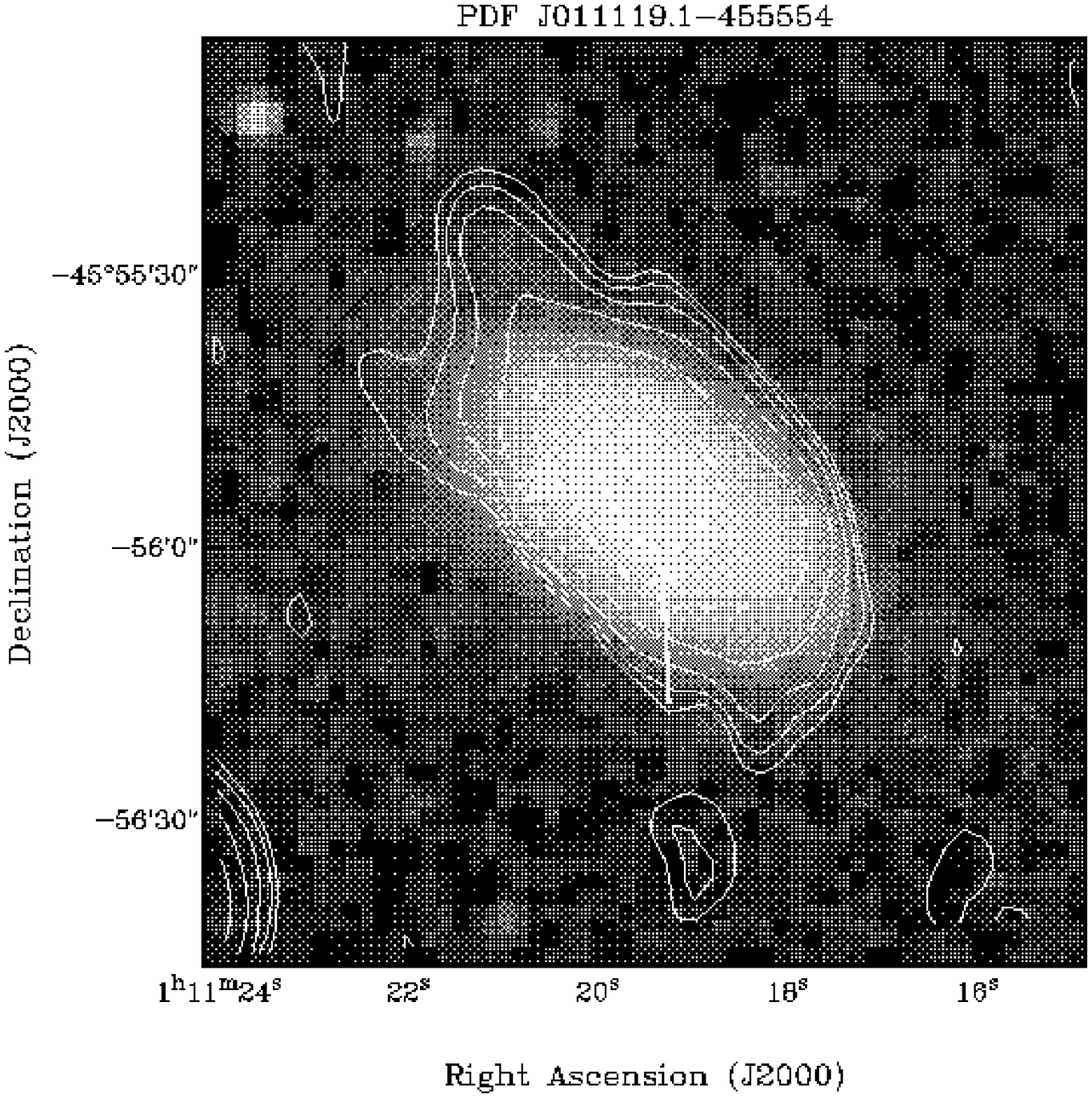}}
\hfill
\rotatebox{0}{\includegraphics[width=5cm]{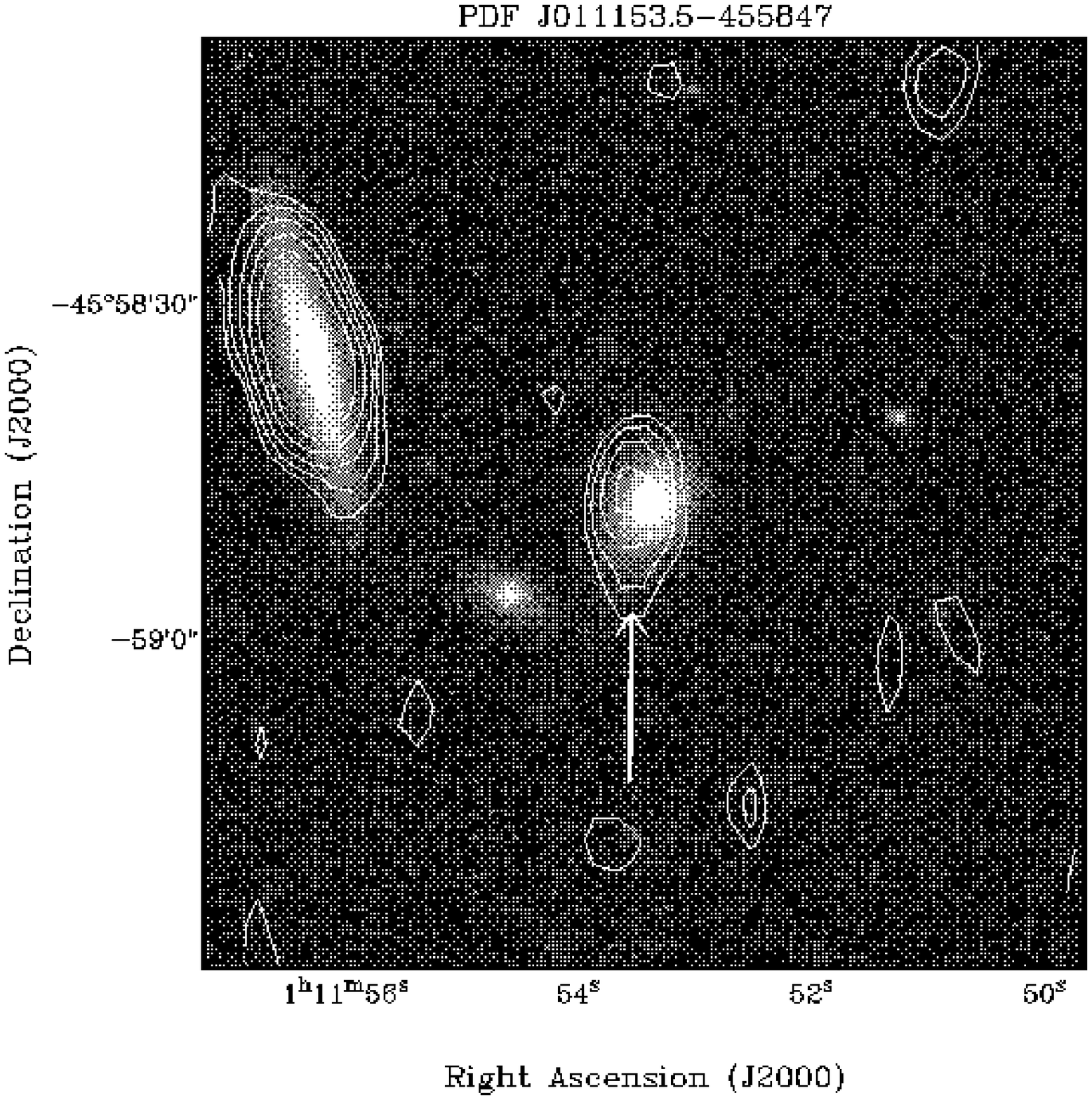}}
\hfill
\rotatebox{0}{\includegraphics[width=5cm]{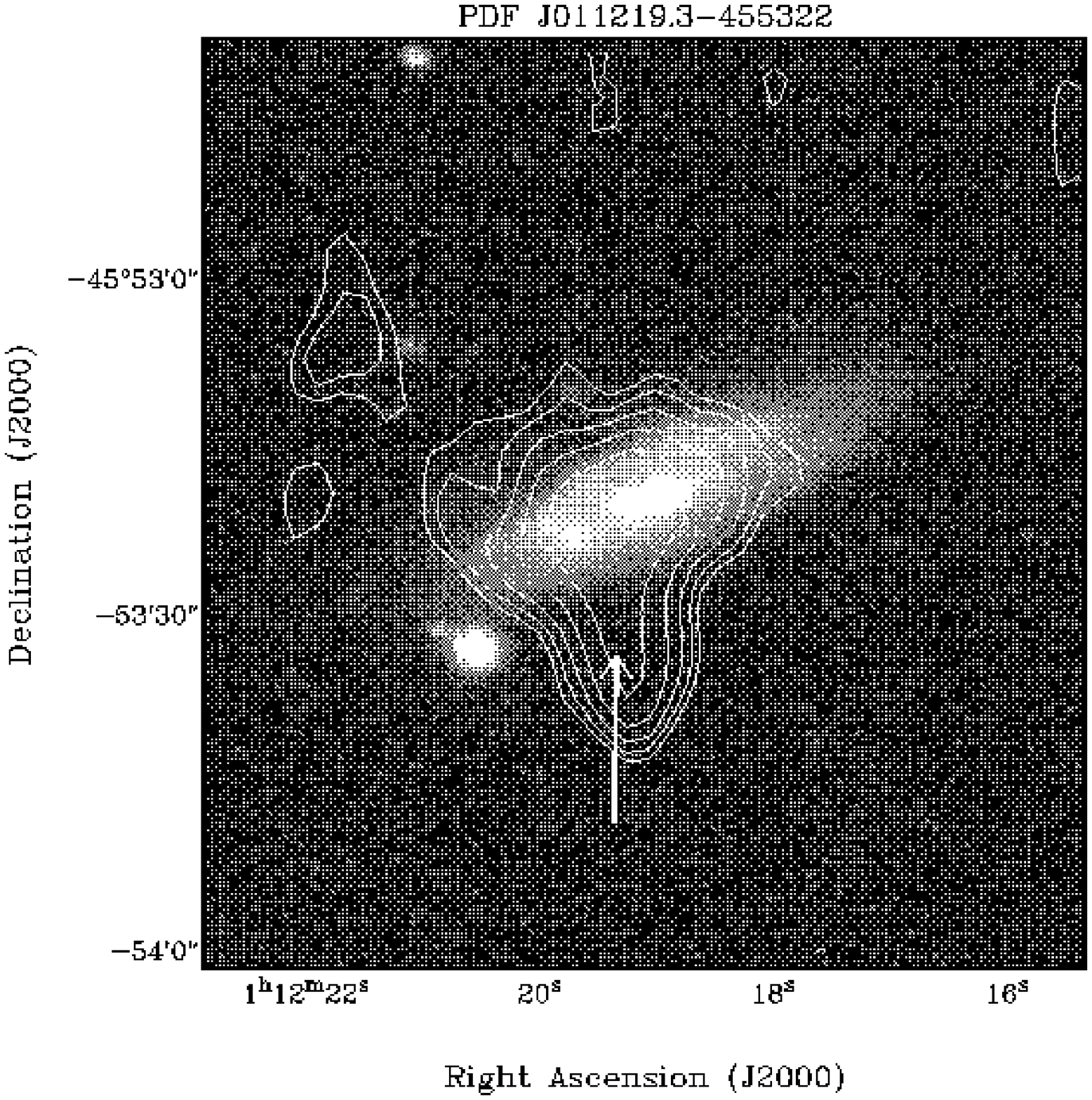}}}
\caption{cont. The left-most image on the bottom row
(PDF~J011119.1$-$455554) is a galaxy which lies outside the region of
the CCD survey and the optical image has been taken from the DSS.
\label{fig1cont}}
\end{figure*}

\begin{figure*}
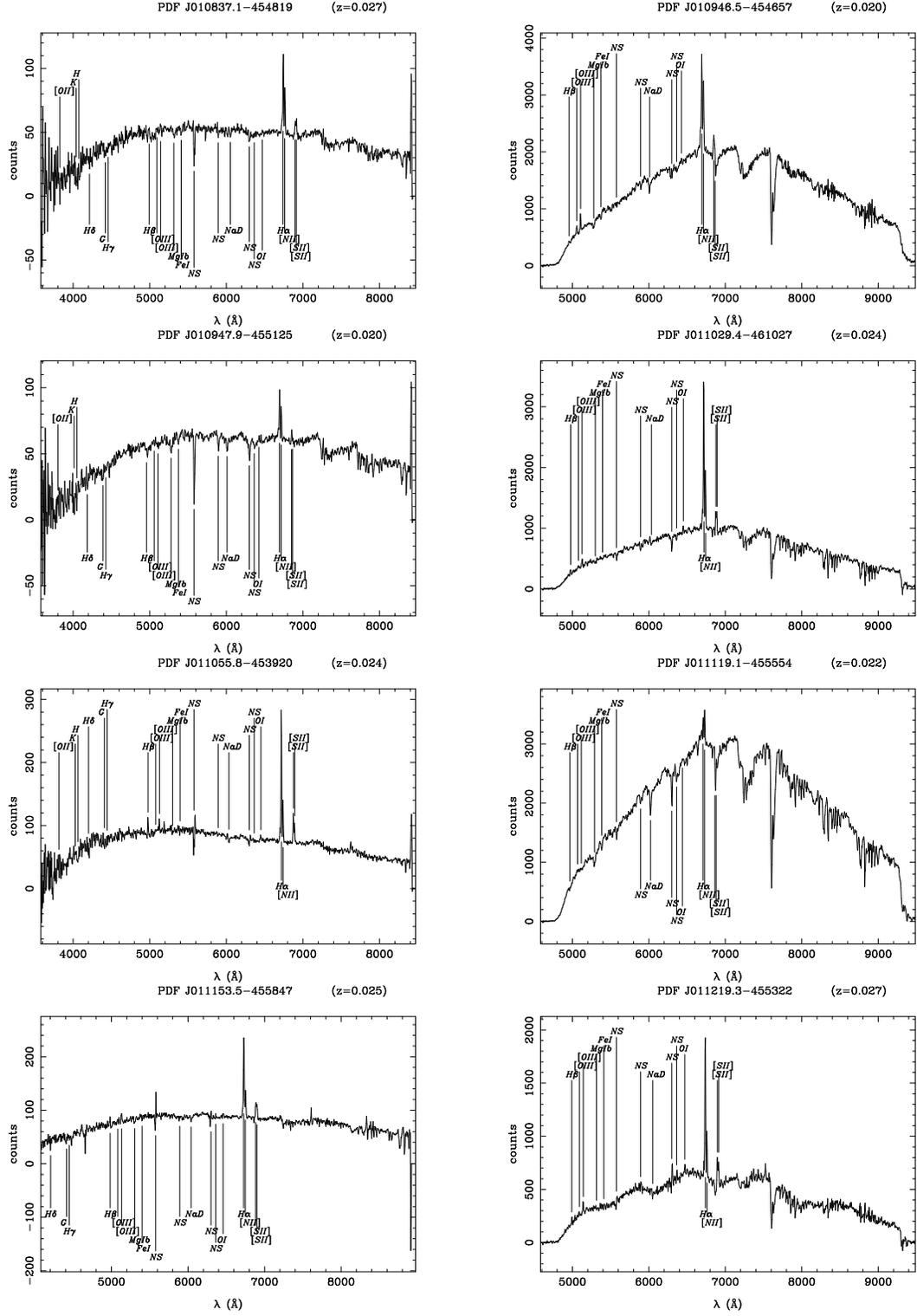

\centerline{\hfill
\rotatebox{-90}{\includegraphics[width=5cm]{spec01.ps}}
\hfill
\rotatebox{-90}{\includegraphics[width=5cm]{spec04.ps}}\hfill}
\centerline{\hfill
\rotatebox{-90}{\includegraphics[width=5cm]{spec05.ps}}
\hfill
\rotatebox{-90}{\includegraphics[width=5cm]{spec10.ps}}\hfill}
\centerline{\hfill
\rotatebox{-90}{\includegraphics[width=5cm]{spec12.ps}}
\hfill
\rotatebox{-90}{\includegraphics[width=5cm]{spec13.ps}}\hfill}
\centerline{\hfill
\rotatebox{-90}{\includegraphics[width=5cm]{spec14.ps}}
\hfill
\rotatebox{-90}{\includegraphics[width=5cm]{spec15.ps}}\hfill}
\caption{Spectra for the eight emission line galaxies. The location of
selected atmospheric emission lines (NS) are marked, as not all have
been successfully removed in each spectrum. \label{spectra}}
\end{figure*}

\begin{figure*}
\centerline{\rotatebox{-90}{\includegraphics[width=7cm]{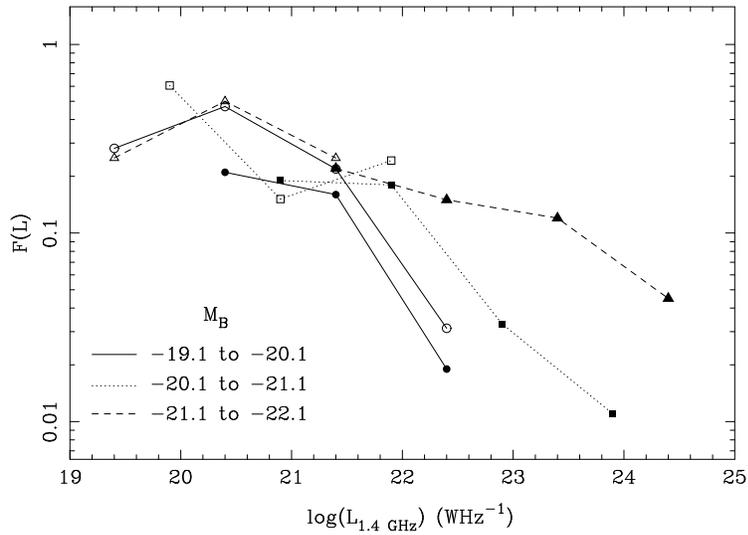}}}
\caption{The FBLF from \protect\citet{Sad:89}, reproduced from
their Figure~5 after converting to $H_0=75\,$kms$^{-1}$Mpc$^{-1}$ and
1.4\,GHz (solid symbols). The open symbols show the
FBLF calculated for A2877. Error bars have been omitted for clarity,
but are of the order of 0.5 dex for the A2877 points.
  \label{flf1}}
\end{figure*}

\end{document}